\documentclass[11pt]{article}
\usepackage{graphicx}
\usepackage{caption}
\usepackage{subcaption}

\usepackage{verbatim}
\usepackage{amsmath}
\usepackage{amssymb}
\usepackage{color}
\usepackage{hyperref}
 \usepackage{stackrel}
 \usepackage{upgreek}
 \usepackage{csquotes}
 \usepackage{fullpage}

 \def\##1{{\bf #1}}
\renewcommand\vec{\mathbf}
\def\doubleunderline#1{\underline{\underline{#1}}}
\def\=#1{\underline{\underline{#1}}}

 \def\eps{\varepsilon}

 \def\epso{\eps_{\scriptscriptstyle 0}}

\def\muo{\mu_{\scriptscriptstyle 0}}
\def\ko{k_{\scriptscriptstyle 0}}
\def\co{c_{\scriptscriptstyle 0}}
\def\koc{{\ko}c}

\def\etao{\eta_{\scriptscriptstyle 0}}

\def\.{\mbox{ \tiny{$^\bullet$} }}

\def\ux{\hat{\#x}}
\def\uy{\hat{\#y}}
\def\uz{\hat{\#z}}
\def\ur{\hat{\#r}}
\def\un{\hat{\#n}}
\def\uphi{\hat{ \boldsymbol\phi}}
\def\utheta{\hat{ \boldsymbol\theta}}
\def\e{\hat{\#e}}
\def\k{\hat{\#k}}

\newenvironment{rcases}
  {\left.\begin{aligned}}
  {\end{aligned}\right\rbrace}

  \def\les{\left[}
\def\ris{\right]}
\def\lec{\left\{}
\def\ric{\right\}}

\def\QD{Q_\text{D}}
\def\Qsca{Q_{\text{sca}}}
\def\Qb{Q_\text{b}}
\def\Qf{Q_\text{f}}

\def\hatGamma{\hat{\#{\Gamma}}}

\def\calV{\cal V}
\def\calS{\cal S}
\def\calVe{{\cal V}_e}
\def\calVi{{\cal V}_i}
\def\calVs{{\cal V}_s}

\begin{document}

\begin{center}
{\sf Scattering by a three-dimensional object composed of the simplest Lorentz-nonreciprocal medium}\\
\vspace{0.5cm}

Hamad M. Alkhoori,$^1$ Akhlesh Lakhtakia,$^2$ James K. Breakall,$^1$ and Craig F. Bohren$^3$\\
\vspace{0.5cm}

$^1${Department of Electrical Engineering, The Pennsylvania State University, University Park, Pennsylvania 16802, USA}\\
$^2${Department of Engineering Science and Mechanics, The Pennsylvania State University, University Park, Pennsylvania 16802, USA}\\
$^3${Department of Meteorology, The Pennsylvania State University, University Park, Pennsylvania 16802, USA} \\
\end{center} 
\vspace{0.5cm}

\begin{abstract}
The simplest  Lorentz-nonreciprocal medium has the constitutive relations (${\#D} =\epso {\#E} -{\#\Gamma}\times {\#H}$ and
${\#B} =\muo {\#H} + {\#\Gamma}\times{\#E}$). Scattering by
 a three-dimensional object composed of this medium was investigated using the extended boundary condition method.
Scattering by this object in free space must be attributed to non-zero $\Gamma=\vert\#\Gamma\vert$.
The differential scattering efficiency is immune to the transformation of the incident toroidal electric field phasor  into a poloidal electric field phasor, or \textit{vice versa}, and  a
consequence of this source-invariance  is the polarization-state invariance of
the  differential scattering efficiency when the irradiating field is a plane wave.  Both the total scattering and forward-scattering efficiencies of  an ellipsoid  composed of the simplest  Lorentz-nonreciprocal medium  are maximum when the plane wave is incident in a direction   {coparallel (but not antiparallel)} to $\#\Gamma$, and the backscattering efficiency is  minimum when $\#\Gamma$  is   {parallel} to the 
incidence  direction. The total scattering and the forward-scattering efficiencies are maximum
when the incidence direction is parallel to the largest semi-axis of the ellipsoid  if the incidence direction is coparallel (but not antiparallel) to $\#\Gamma$.
Lorentz nonreciprocity in an object is thus intimately connected to the shape of that object in affecting the scattered field.

\end{abstract}

\def\doubleunderline#1{\underline{\underline{#1}}}
\renewcommand\vec{\mathbf}
\section{Introduction} \label{s1}

A remarkable feature of 21st-century electromagnetics is the diversity of linear materials in which electromagnetic phenomena
are being investigated. These materials can be isotropic, biisotropic, anisotropic or bianisotropic \cite{Nye,Charney,EAB, Kong}. These materials can be reciprocal
or nonreciprocal in the Lorentz sense \cite{Krowne}. Natural examples of some types of these materials may not be known or are very uncommon,
but can be engineered\cite{Marques,Cai,Smoly}  as composite materials \cite{Neelakanta,MAEH}. 

Several composite materials have been proposed to  replicate certain characteristics of relativistic spacetime \cite{Li1,Li2,ML1,ML2,Li3,ML3,Khvesh}, based on a theorem of Pl\'ebanski \cite{Plebanski} according to which relativistic spacetime can be replaced by a bianisotropic continuum. This continuum
may exhibit the magnetoelectric \cite{Odell} effect in a Lorentz-nonreciprocal sense \cite{ML-njp}. Suppose that relativistic spacetime
is described by the gravitational metric $g_{\alpha\beta}$, $\alpha\in\lec0,1,2,3\ric$
and $\beta\in\lec0,1,2,3\ric$, with $(+,-,-,-)$ as its  signature. We can then define a dyadic $\=\gamma$ 
and a vector $\#\Gamma$ with components $\gamma_{\ell{m}}=-({-\bar{g}})^{1/2}g^{\ell{m}}/g_{00}$
and $\Gamma_\ell=\co^{-1}g_{0\ell}/g_{00}$,  $\ell\in\lec1,2,3\ric$
and $m\in\lec1,2,3\ric$, where $\bar{g}$ denotes the determinant of  $g_{\alpha\beta}$,
$\co=1/\sqrt{\epso\muo}$, and $\epso$ and $\muo$ are the permittivity and the permeability of 
the gravitationally unaffected   free space, respectively.
 According to the Pl\'ebanski
theorem, the constitutive equations of the equivalent bianisotropic continuum are
\begin{equation}
\label{conrel0}
\left.\begin{array}{l}
{\#D} =\epso\=\gamma\.{\#E} -\left({\#\Gamma}\times\=I\right)\. {\#H}
\\[5pt]
{\#B} =\muo\=\gamma\.{\#H} + \left({\#\Gamma}\times\=I\right)\.{\#E}
\end{array}\right\}\,,
\end{equation}
where $\=I$ is the identity dyadic \cite{Chen}.

The vector $\#\Gamma$ may be called the magnetoelectric-gyrotropy vector. It has the same role as   a bias field
in a magnetoplasma or a ferrite \cite{Felsen,Chen,Inan}; hence the use of the term \textit{gyrotropy}. However, 
the Pl\'ebanski medium specified by Eq.~(\ref{conrel0}) is distinct from  magnetoplasmas and ferrites since the gyrotropy in the former case is
present in the magnetoelectric dyadics, whereas it is present in the
permittivity and permeability dyadics in the latter cases.

Lorentz nonreciprocity is signalled by $\Gamma=\vert\#\Gamma\vert\ne0$,  because $\#\Gamma \times \={I}$ is an antisymmetric dyadic \cite{Krowne}.  
In contrast,  gravitationally unaffected free space is isotropic and Lorentz reciprocal. Thus, a Lorentz-nonreciprocal counterpart of free space emerges from the spacetime metric
\cite{Jafri2}
\begin{eqnarray}
\les{g_{\alpha\beta}}\ris&=& u^{-1}\left(u+w_1^2+w_2^2+w_3^2\right)^{-1/4}
\les
\begin{array}{cccc}
1 & w_1 & w_2 & w_3\\
w_1 & -u & 0 & 0\\
w_2 & 0 &-u & 0\\
w_3 & 0 & 0 & -u
\end{array}\ris\,,
\label{metric1}
\end{eqnarray}
where $w_{1}$, $w_{2}$, and $w_{3}$  are the real-valued components of a vector $\#w$ and $u>0$.
Then, $\bar{g}u^2=-1$ and   Eqs.~(\ref{conrel0}) hold with 
$\#\Gamma=\co^{-1}\#w$ and $\=\gamma = \=I$.
The constitutive relations of this Lorentz-nonreciprocal medium thus are as follows:
\begin{equation}
\label{conrel1}
\left.\begin{array}{l}
{\#D} =\epso {\#E} -\left({\#\Gamma}\times\=I\right)\. {\#H}
\\[5pt]
{\#B} =\muo {\#H} + \left({\#\Gamma}\times\=I\right)\.{\#E}
\end{array}\right\}\,.
\end{equation}
This medium   is the \textit{simplest} Lorentz-nonreciprocal medium 
possible because it differs from     free space (${\#D} =\epso {\#E}$
and ${\#B} =\muo {\#H}$), the reference medium
in electromagnetic theory, by
a single constitutive scalar: $\Gamma$.
A bounded object composed of  the simplest Lorentz-nonreciprocal medium and
suspended in     free space can scatter only 
if $\Gamma\ne0$.

As the simplest Lorentz-nonreciprocal medium can be transformed into    free space by a straightforward transformation
of electric and magnetic field phasors \cite{LW1997,DGFbook}, simple solutions of the frequency-domain Maxwell
equations with Eqs.~(\ref{conrel1}) substituted exist. Accordingly,
plane-wave scattering by a sphere composed of the simplest Lorentz-nonreciprocal medium 
has been theoretically examined \cite{Jafri2}. The scattering characteristics depend 
strongly on the magnitude and direction of $\#\Gamma$. The total scattering and forward-scattering efficiencies are more pronounced when $\#\Gamma$ is parallel to the propagation direction of the incident plane wave than when it is parallel to either the
incident electric field phasor or the incident magnetic field phasor. Also, the backscattering efficiency vanishes when $\#\Gamma$ is parallel to the propagation direction of the incident plane wave. 
But because the scattering object is a sphere, the role of the shape of the object on scattering characteristics
could not be unveiled.

Here, we investigate the simultaneous effects of shape and magnetoelectric  gyrotropy
 by formulating and solving the problem of plane-wave scattering by  an object composed of the simplest Lorentz-nonreciprocal medium
 suspended in   free space. The extended boundary condition method (EBCM), also called the null-field method and the T-matrix method \cite{Waterman}, was adopted to solve the scattering problem. The scattered and internal field phasors were expanded in terms of appropriate vector spherical wavefunctions \cite{Waterman} with  unknown expansion coefficients, and  the  incident field phasors were similarly expanded but with  known expansion coefficients. Application of the Ewald--Oseen extinction theorem 
 and the Huygens principle  then yielded  a transition matrix  to relate the scattered-field coefficients to the incident-field coefficients \cite{Lakh-IskBook}.

The plan of the paper is as follows. The EBCM  equations for the chosen scattering problem are presented in Sec.~\ref{s2}.
These equations are exploited in Sec.~\ref{sourceinv} to uncover a source-invariance property of the differential scattering efficiency.
Furthermore, we show in Sec.~\ref{sourceinv-plw} that the plane-wave scattering efficiencies are not affected by the
polarization state of the incident plane wave.
The  plane-wave scattering efficiencies of an ellipsoid composed
of the simplest Lorentz-nonreciprocal medium
 are presented in Sec.~\ref{s3}
in relation to the direction of propagation and the polarization state of the incident plane wave, the shape of the ellipsoid, and the 
magnetoelectric-gyrotropy vector.  Conclusions are summarized in Sec.~\ref{s4}. 

An $\exp(-i \omega t)$ dependence on time $t$ is implicit  throughout the analysis with $i=\sqrt{-1}$ and
$\omega$ is the angular frequency. The wavenumber in   free space is denoted by $\ko=\omega/\co$ and the
intrinsic impedance of    free space by $\etao=\sqrt{\muo/\epso}$. 
Vectors are  in boldface, unit vectors are decorated by carets,  dyadics are  double underlined, and column vectors as well as matrices are enclosed in square brackets.

\section{EBCM Equations}\label{s2} 
Let all space $\calV$ be partitioned into   into two
  disjoint regions $\calVe$ and $\calVi$ separated by the closed surface $\calS$.
Extending to infinity in all directions, the exterior region $\calVe$ is vacuous except that a portion $\calVs\subset\calVe$
is occupied by the source of a time-harmonic electromagnetic field. The interior region $\calVi$ is
occupied by the the simplest Lorentz-nonreciprocal medium. The origin of the coordinate system lies in $\calVi$.

Application of the Ewald--Oseen extinction theorem and the Huygens principle leads to
the integral equations \cite{Waterman,Lakh-IskBook}
\begin{eqnarray}
\nonumber
&&
\displaystyle{
\left.\begin{array}{c}
\#E_{\text{sca}}(\#r) 
\\[5pt]
-\#E_{\text{inc}}(\#r) 
\end{array}\right\}
=
\iint\limits_{\calS}\lec-\=G^{em}(\#r,\#r_s)\.
\les\un(\#r_s)\times\#E_{int}(\#r_s)\ris \right.
}
\\[10pt]
&&\,  
\displaystyle{
\left. +\=G^{ee}(\#r,\#r_s)\.
\les\un(\#r_s)\times\#H_{\text{int}}(\#r_s)\ris\ric
d^2r_s\,,\quad
\left\{\begin{array}{l}
\#r\in\calVe
\\[5pt]
\#r\in\calVi
\end{array}\right..
}
\label{newE74}
\end{eqnarray}
In these equations, $\#E_{\text{inc}}(\#r)$ is the incident electric field phasor, $\#E_{\text{sca}}(\#r)$
is the scattered electric field phasor,  $\#E_{\text{int}}(\#r)$ is the internal electric field phasor, 
  $\#H_{\text{int}}(\#r)$ is the internal magnetic field phasor, $\un(\#r_s)$ is the unit outward
  normal to $\calS$ at $\#r_s\in\calS$, 
\begin{equation}
\=G^{em}(\#r,\#r_s) =-\frac{1} {i\omega\muo }\nabla\times\=G^{ee}(\#r,\#r_s)\,,
\end{equation}
and
\begin{equation}
\=G^{ee}(\#r,\#r_s) = i\omega\muo \left( \=I + \frac{\nabla\nabla}{\ko^2}\right)
\frac{\exp\left(i\ko\vert\#r-\#r_s\vert\right)}
{4\pi\vert\#r-\#r_s\vert}
\end{equation}
is the dyadic Green function for free space \cite{Chen}.

\subsection{Incident electric and magnetic field phasors}
The incident electric  field phasor is expressed as
\begin{equation}
 \label{1}
\begin{aligned}
\vec{E}_{\text{inc}}(\vec{r}) =&  \lim_{N\to\infty} \sum_{s \in \{e,o\}} \sum_{n=1}^N \sum_{m=0}^n  \big\{
D_{mn}\big[A_{smn}^{(1)} \vec{M}_{smn}^{(1)}(\ko \vec{r}) \\
& + B_{smn}^{(1)} \vec{N}_{smn}^{(1)}(\ko \vec{r}) \big]\big\}\,
\end{aligned}
\end{equation}
and the incident magnetic field phasor as
\begin{equation}
 \label{1-H}
\begin{aligned}
\vec{H}_{\text{inc}}(\vec{r}) =&-\frac{i}{\etao}  \lim_{N\to\infty} \sum_{s \in \{e,o\}} \sum_{n=1}^N \sum_{m=0}^n 
 \big\{ D_{mn}\big[ A_{smn}^{(1)} \vec{N}_{smn}^{(1)}(\ko \vec{r}) \\
& + B_{smn}^{(1)} \vec{M}_{smn}^{(1)}(\ko \vec{r}) \big]\big\}\,,
\end{aligned}
\end{equation}
where the normalization factor
\begin{equation}
D_{mn} = (2- \delta_{m0}) \frac{(2n+1)(n-m)!}{4n(n+1)(n+m)!}
\end{equation}
involves the Kronecker delta $\delta_{m m^\prime}$. 

The vector spherical wavefunctions of the first kind,
$\vec{M}_{smn}^{(1)}(\ko \vec{r})$ and $\vec{N}_{smn}^{(1)}(\ko \vec{r})$, are  available in standard texts \cite{Morse,Stratton,Collin}, 
the  index $n$ denoting the order of the spherical Bessel function $j_n(\ko{r})$ appearing in those wavefunctions.
The index $n$ is restricted to $[1,N]$ where $N$ is sufficiently large for acceptable convergence  
and the limit on the right sides of Eqs.~(\ref{1}) and (\ref{1-H})  is not used.  The vector spherical wavefunctions also contain
the associated Legendre function $P_n^m(\cos \theta)$    of order $n$ and degree $m$,
and the index $s$ stands for either  even (e) or odd (o) parity. 

The  column vectors $[A^{(1)}]$ and $[B^{(1)}]$ of the
expansion coefficients $A_{smn}^{(1)}$ and $B_{smn}^{(1)}$, respectively, are assumed to be known. An assumption underlying
Eqs.~(\ref{1}) and (\ref{1-H}) is that the source of the incident field phasors is not affected by scattering by the object.

\subsection{Scattered electric and magnetic field phasors}

 The scattered electric and magnetic field phasors take the form
 \begin{equation}
 \label{2}
\begin{aligned}
\vec{E}_{\text{sca}}(\vec{r}) =&  \lim_{N\to\infty} \sum_{s \in \{e,o\}} \sum_{n=1}^N \sum_{m=0}^n
\big\{  D_{mn}\big[ A_{smn}^{(3)} \vec{M}_{smn}^{({3})}(\ko \vec{r}) \\
& + B_{smn}^{(3)} \vec{N}_{smn}^{({3})}(\ko \vec{r}) \big]\big\}\,
\end{aligned}
\end{equation}
and 
 \begin{equation}
 \label{2-H}
\begin{aligned}
\vec{H}_{\text{sca}}(\vec{r}) =&-\frac{i}{\etao}  \lim_{N\to\infty} \sum_{s \in \{e,o\}} \sum_{n=1}^N \sum_{m=0}^n
\big\{  D_{mn}\big[ A_{smn}^{(3)} \vec{N}_{smn}^{({3})}(\ko \vec{r}) \\
& + B_{smn}^{(3)} \vec{M}_{smn}^{({3})}(\ko \vec{r}) \big]\big\}\,.
\end{aligned}
\end{equation}
 
The vector spherical wavefunctions of the third kind \cite{Morse,Stratton, Collin}, $\vec{M}_{smn}^{(3)}(\ko \vec{r}) $ and $\vec{N}_{smn}^{(3)}(\ko \vec{r}) $, involve the spherical Hankel function $h_n^{(1)}(\ko{r})$ instead of $j_n(\ko{r})$. The column vectors $[A^{(3)}]$ and $[B^{(3)}]$ of the
 expansion coefficients $A_{smn}^{(3)}$ and $B_{smn}^{(3)}$, respectively, are unknown. In a strict sense, Eqs.~(\ref{2}) and (\ref{2-H}) hold outside the smallest sphere circumscribing $\calVi$ with its center on the
origin of the coordinate system \cite{Lakh-IskBook}.

The scattered electric field phasor
can be approximated as \cite{Bowman,Bohren}
\begin{equation}
\vec{E}_{\text{sca}}(r,\theta,\phi) \approx \vec{F}_{\text{sca}}( \theta,\phi) \frac{\exp(i\ko r)}{r}
\end{equation}
as ${\ko}r\to\infty$,
where \cite{Alkhoori1}
\begin{equation}
\begin{aligned}[b]
\vec{F}_{\text{sca}}(\theta,\phi)=& \frac{1}{\ko}\lim_{N\to\infty}
 \sum_{s \in \{e,o\}} \sum_{n=1}^{N} \sum_{m=0}^n \bigg\{(-i)^n D_{mn} \sqrt{n(n+1)} \\
& \big[ -i A_{smn}^{(3)} \vec{C}_{smn}(\theta,\phi) + B_{smn}^{(3)}\ur \times
\vec{C}_{smn}(\theta, \phi) \big] \bigg\}
\end{aligned}
\end{equation}
involves
\begin{eqnarray}
\nonumber
&&
 \vec{C}_{smn}(\theta,\phi)=
  \frac{1}{\sqrt{n(n+1)}} \left[ \mp m \frac{P_n^m(\cos \theta)}{\sin \theta}  \Bigg\{ \begin{matrix}
\sin (m \phi)\\
\cos (m \phi)
\end{matrix} \Bigg\} \utheta \right.
\\[8pt]
&& \quad\left.-\frac{d P_n^m(\cos \theta)}{d \theta} \Bigg\{ \begin{matrix}
\cos (m\phi)\\
\sin (m\phi)
\end{matrix} \Bigg\}  \uphi  \right]  \,,\quad
s=\left\{\begin{array}{l} e \\ o\end{array}\right.\,.
\end{eqnarray}

The differential scattering efficiency is defined as
\begin{equation} \label{QD-def}
\QD(\theta,\phi)= \frac{4}{c^2}\,  \vert\vec{F}_{\text{sca}}(\theta,\phi)\vert^2 \,,
\end{equation}
where $c$ is an appropriate linear dimension of the object.
The total scattering efficiency is
obtained as
\begin{equation} \label{Qsca-def}
\Qsca=  \frac{1}{\left(\koc\right)^2}\lim_{N\to\infty} \sum_{s \in \{e,o\}} \sum_{n=1}^{N} \sum_{m=0}^n  D_{mn} \left[ \vert{A_{smn}^{(3)}}\vert^2+ \vert{B_{smn}^{(3)}}\vert^2 \right]\,.
\end{equation}

\subsection{Internal electric and magnetic field phasors}
The electric and magnetic field phasors excited inside the scattering object are represented by \cite{LW1997,DGFbook}
\begin{equation} \label{Eint}
\begin{aligned}
\vec{E}_{\text{int}}(\vec{r}) =&   \exp(i\omega \#\Gamma\.\vec{r}) \lim_{N\to\infty} \sum_{s \in \{e,o\}} \sum_{n=1}^N \sum_{m=0}^n  [ b_{smn} \vec{M}_{smn}^{(1)}( {\ko}\vec{r}) \\
 &+ c_{smn} \vec{N}_{smn}^{(1)}( {\ko}\vec{r})  \big],
\end{aligned}
\end{equation}
and
\begin{equation} \label{Hint}
\begin{aligned}
\vec{H}_{\text{int}}(\vec{r}) =&  -\frac{i}{\etao} \exp(i\omega \#{\Gamma}\.\vec{r}) \lim_{N\to\infty} \sum_{s \in \{e,o\}} \sum_{n=1}^N \sum_{m=0}^n  [ b_{smn} \vec{N}_{smn}^{(1)}( {\ko}\vec{r}) \\
 &+ c_{smn} \vec{M}_{smn}^{(1)}( {\ko}\vec{r})  \big],
\end{aligned}
\end{equation}
where the    column vectors $[b]$ and $[c]$ comprising the expansion coefficients $b_{smn}$ and $c_{smn}$,
respectively,  have to be determined.

\subsection{Transition matrix}

Equations~(\ref{1}), (\ref{1-H}), (\ref{2}), (\ref{2-H}), (\ref{Eint}), and (\ref{Hint}) are substituted in
Eqs.~(\ref{newE74}) along with the bilinear expansion \cite{DGFbook} of $\=G^{ee}(\#r,\#r_s)$ in terms of the
vector spherical wavefunctions. After using the
 the orthogonality properties of
the vector spherical wavefunctions on unit spheres \cite{Stratton}, the incident-field coefficients
and the scattered-field coefficients can be related to the tangential components of the electric and magnetic field phasors on 
the exterior side of $S$. Standard boundary conditions then connect those tangential components to the internal
electric and magnetic field phasors evaluated on the interior side of $S$ \cite{Lakh-IskBook}.

A set of algebraic equations thereby emerges to relate
the incident-field coefficients to the internal-field coefficients \cite{Waterman}:
\begin{equation}
\begin{bmatrix}
\les{A^{(1)}}\ris \\ --- \\ \les{ B^{(1)}}\ris
\end{bmatrix}
=[Y^{(1)}]
\begin{bmatrix}
\les{b} \ris \\ --- \\  \les{c}\ris
\end{bmatrix}\,,
\label{eq19}
\end{equation} 
in matrix notation. Similarly, a set of algebraic equations  emerges to relate
the scattered-field coefficients to the internal-field coefficients as follows:
\begin{equation}
\begin{bmatrix}
\les{A^{(3)}}\ris \\ --- \\ \les{ B^{(3)}}\ris
\end{bmatrix}
=-[Y^{(1)}]
\begin{bmatrix}
\les{b} \ris \\ --- \\  \les{c}\ris
\end{bmatrix}\,.
\label{eq20}
\end{equation} 
Therefore,
\begin{equation}
\begin{bmatrix}
\les{A^{(3)}}\ris \\ --- \\ \les{ B^{(3)}}\ris
\end{bmatrix}
=[T]
\begin{bmatrix}
\les{A^{(1)}}\ris \\ --- \\ \les{ B^{(1)}}\ris
\end{bmatrix}\,,
\label{eq21}
\end{equation} 
where 
\begin{equation} 
\label{4a}
[T]= -[Y^{(3)}] [Y^{(1)}]^{-1}
\end{equation}
is the transition matrix.

The matrix $[Y^{(j)}]$, $j\in[1,3]$, is symbolically written as
\begin{equation} 
\label{23}
[Y^{(j)}]= 
\begin{pmatrix}
I_{smn,s^\prime m^\prime n^\prime}^{(j)} && \big| && J_{smn,s^\prime m^\prime n^\prime}^{(j)} \\
---- && \big| && ---- \\
J_{smn,s^\prime m^\prime n^\prime}^{(j)}&& \big| && I_{smn,s^\prime m^\prime n^\prime}^{(j)}
\end{pmatrix}.
\end{equation}
The matrix elements in Eq. (\ref{23}) are surface integrals given by 
 \begin{eqnarray} 
\nonumber
I_{smn,s^\prime m^\prime n^\prime}^{(j)} &=&
- \frac{i {\ko}^2}{\pi}
\iint\limits_{\calS}
\Big\{ \vec{N}_{smn}^{(\ell)}({\ko} \vec{r}_s) \. [\un(\vec{r}_s)   \times \vec{M}^{(1)}_{s^\prime m^\prime n^\prime}({\ko} \vec{r}_s)]   \\
\nonumber
&&\quad+ \vec{M}_{smn}^{(\ell)}({\ko} \vec{r}_s) \. [\un(\vec{r}_s)   \times \vec{N}^{(1)}_{s^\prime m^\prime n^\prime}({\ko} \vec{r}_s)]\Big\} 
\\
&&\qquad\times\exp(i\omega \#\Gamma\.\vec{r}_s) \,d^2r_s
\end{eqnarray}
and
 \begin{eqnarray} 
\nonumber
J_{smn,s^\prime m^\prime n^\prime}^{(j)} &=&
- \frac{i {\ko}^2}{\pi}
\iint\limits_{\calS}
\Big\{ \vec{N}_{smn}^{(\ell)}({\ko} \vec{r}_s) \. [\un(\vec{r}_s)   \times \vec{N}^{(1)}_{s^\prime m^\prime n^\prime}({\ko} \vec{r}_s)]   \\
\nonumber
&&\quad+ \vec{M}_{smn}^{(\ell)}({\ko} \vec{r}_s) \. [\un(\vec{r}_s)   \times \vec{M}^{(1)}_{s^\prime m^\prime n^\prime}({\ko} \vec{r}_s)]\Big\} 
\\
&&\qquad\times\exp(i\omega \#\Gamma\.\vec{r}_s) \,d^2r_s\,.
\end{eqnarray}
The symmetries
\begin{equation}
\left.\begin{array}{l}
I_{smn,s^\prime m^\prime n^\prime}^{(j)}=-I_{s^\prime m^\prime n^\prime,smn}^{(j)}
\\
J_{smn,s^\prime m^\prime n^\prime}^{(j)}=-J_{s^\prime m^\prime n^\prime,smn}^{(j)}
\end{array}\right\}
\end{equation}
should be used to  increase computational speed.

\section{Toroidal-poloidal source invariance of differential scattering efficiency} \label{sourceinv}
A remarkable feature of an object composed of
the simplest Lorentz-nonreciprocal medium becomes evident on exchanging
the  column vectors $[A^{(1)}]$ and $[B^{(1)}]$ on the left side of Eq.~(\ref{eq19}). This equation remains
invariant if the  column vectors $[b]$ and $[c]$ on its right side are   interchanged as well. If
the  column vectors $[A^{(3)}]$ and $[B^{(3)}]$ on the left side of Eq.~(\ref{eq20}) are also interchanged
at the same time, then Eqs.~(\ref{eq20}) and (\ref{eq21}) also remain unchanged.

This property leads to the toroidal-poloidal source invariance of the differential scattering efficiency of any
object composed of
the simplest Lorentz-nonreciprocal medium as follows. Let the incident electric
field phasors from sources I and II be given by
\begin{equation}
 \label{1-I}
\begin{aligned}
\vec{E}_{\text{inc}}^{I}(\vec{r}) =&  \lim_{N\to\infty} \sum_{s \in \{e,o\}} \sum_{n=1}^N \sum_{m=0}^n  \big\{
D_{mn}\big[A_{smn}^{(1)} \vec{M}_{smn}^{(1)}(\ko \vec{r}) \\
& + B_{smn}^{(1)} \vec{N}_{smn}^{(1)}(\ko \vec{r}) \big]\big\}\,
\end{aligned}
\end{equation}
and
\begin{equation}
 \label{1-II}
\begin{aligned}
\vec{E}_{\text{inc}}^{II}(\vec{r}) =&  \lim_{N\to\infty} \sum_{s \in \{e,o\}} \sum_{n=1}^N \sum_{m=0}^n  \big\{
D_{mn}\big[B_{smn}^{(1)} \vec{M}_{smn}^{(1)}(\ko \vec{r}) \\
& + A_{smn}^{(1)} \vec{N}_{smn}^{(1)}(\ko \vec{r}) \big]\big\}\,
\end{aligned}
\end{equation}
  Then, the corresponding scattered electric field phasors must be
\begin{equation}
 \label{sca-I}
\begin{aligned}
\vec{E}_{\text{sca}}^{I}(\vec{r}) =&  \lim_{N\to\infty} \sum_{s \in \{e,o\}} \sum_{n=1}^N \sum_{m=0}^n  \big\{
D_{mn}\big[A_{smn}^{(3)} \vec{M}_{smn}^{(3)}(\ko \vec{r}) \\
& + B_{smn}^{(3)} \vec{N}_{smn}^{(3)}(\ko \vec{r}) \big]\big\}\,
\end{aligned}
\end{equation}
and
\begin{equation}
 \label{sca-II}
\begin{aligned}
\vec{E}_{\text{sca}}^{II}(\vec{r}) =&  \lim_{N\to\infty} \sum_{s \in \{e,o\}} \sum_{n=1}^N \sum_{m=0}^n  \big\{
D_{mn}\big[B_{smn}^{(3)} \vec{M}_{smn}^{(3)}(\ko \vec{r}) \\
& + A_{smn}^{(3)} \vec{N}_{smn}^{(3)}(\ko \vec{r}) \big]\big\}\,.
\end{aligned}
\end{equation}
In consequence, after noting that $\ur\.\#C_{smn}(\theta,\phi)\equiv 0$, we see that
\begin{equation}
\label{eq31}
\QD^{I}(\theta,\phi)=\QD^{II}(\theta,\phi)\,.
\end{equation}

The vector spherical wave functions $\#M_{smn}^{(j)}$ and $\#N_{smn}^{(j)}$ are toroidal and poloidal, respectively
\cite{CK1957, DS1993}. The curl of a toroidal/poloidal function is poloidal/toroidal. Accordingly, the source of
a toroidal/poloidal electric field phasor is also the source of a poloidal/toroidal magnetic field phasor. Equation~(\ref{eq31})
demonstrates that the transformation of a source of a toroidal electric field phasor to the source of a poloidal
electric field phasor, or \textit{vice versa}, does not affect the differential scattering efficiency of an object composed of
the simplest Lorentz-nonreciprocal medium.
This toroidal-poloidal source invariance  extends to the total scattering efficiency
$\Qsca$.

\section{Polarization-state invariance of plane-wave scattering efficiencies} \label{sourceinv-plw}
As a corollary of the  toroidal-poloidal source invariance of the differential scattering efficiency of any
object composed of
the simplest Lorentz-nonreciprocal medium, the polarization-state invariance holds for the plane-wave scattering efficiencies of that object. The electric    field phasor of an incident plane wave is given as
\begin{equation}
\label{Einc}
\vec{E}_{\text{inc}}(\vec{r})= \e_{\text{inc}} \exp(i\vec{k}_{\text{inc}} \. \vec{r})\,,
\end{equation}
where  the
unit vector $\e_{\text{inc}}$ defines the polarization state and
the wave vector
\begin{equation}
\vec{k}_{\text{inc}}= \ko(\ux \sin \theta_{\text{inc}} \cos \phi_{\text{inc}} + \uy \sin \theta_{\text{inc}} \sin \phi_{\text{inc}}+\uz \cos \theta_{\text{inc}}),
\end{equation}
involves the angles $\theta_{\text{inc}}\in[0,\pi]$ and $\phi_{\text{inc}}\in[0,2\pi)$ that define the incidence direction.  
The corresponding magnetic field phasor is given as
\begin{equation}
\vec{H}_{\text{inc}}(\vec{r})=\frac{\vec{k}_{\text{inc}}\times\e_{\text{inc}}}{\omega\muo}\, \exp\left(i\vec{k}_{\text{inc}}\.\#r\right)\,.
\end{equation}
The unit vector $\k_{\text{inc}}=\vec{k}_{\text{inc}}/ \ko$ is defined for later convenience and is to be
understood as equivalent to the pair $\lec\theta_{\text{inc}},\phi_{\text{inc}}\ric$. The expansion coefficients  are given by \cite{Morse,Alkhoori1}
\begin{equation}
\begin{rcases}
\begin{aligned}[b]
A_{smn}^{(1)} = 4 i^n \sqrt{n(n+1)}\, \e_{\text{inc}} \. \vec{C}_{smn}(\theta_{\text{inc}},\phi_{\text{inc}})  \\
B_{smn}^{(1)} = 4 i^{n-1} \sqrt{n(n+1)}\, \e_{\text{inc}} \.  
\les\k_{\text{inc}}\times
\vec{C}_{smn}(\theta_{\text{inc}},\phi_{\text{inc}}) \ris
\end{aligned}
\end{rcases}\,,
\end{equation}

For a fixed $\k_{\text{inc}}$, $\QD(\theta,\phi)$ is invariant under a $90^\circ$ rotation of $\e_{\text{inc}}$ about $\k_{\text{inc}}$,  independently of
 the shape of the object composed of
the simplest Lorentz-nonreciprocal medium.
This feature can be derived as follows. 
\begin{itemize}
\item A $90^\circ$ rotation of $\e_{\text{inc}}$ about $\k_{\text{inc}}$ results in the transformation 
\begin{equation} \label{incident-trans}
 \lec A_{smn}^{(1)},B_{smn}^{(1)}\ric \to \lec -iB_{smn}^{(1)},-iA_{smn}^{(1)}\ric\,. \end{equation}
\item Equation~(\ref{eq19}) then remains unchanged if the transformation 
\begin{equation} 
\label{internal-trans}
 \lec b_{smn} ,c_{smn} \ric \to
\lec -ic_{smn},-ib_{smn}\ric\,
\end{equation}
is carried out.
\item Equation~(\ref{eq20}) then remains unchanged if the transformation 
\begin{equation} 
\label{scattered-trans}
  \lec A_{smn}^{(3)},B_{smn}^{(3)}\ric \to
\lec -iB_{smn}^{(3)},-iA_{smn}^{(3)}\ric\,
\end{equation}
is implemented.
\item Use of the transformation of Eq.~(\ref{scattered-trans}) in Eq. (\ref{QD-def}) shows that 
$ \QD(\theta,\phi)$ remains unchanged.
\end{itemize}
The invariances of  the forward-scattering efficiency
\begin{equation} 
\label{Qf-def}
\Qf= \QD(\k_{\text{inc}})\,
\end{equation}
and the backscattering efficiency
\begin{equation} 
\label{Qb-def}
\Qb= \QD(-\k_{\text{inc}})\,
\end{equation}
follow from the invariance of $\QD(\theta,\phi)$, as also does the invariance of
$\Qsca$.

Thus,  all scattering efficiencies of an object composed of
the simplest Lorentz-nonreciprocal medium  will not be affected at all by the polarization state of the incident plane wave,  
regardless of the shape of the object. 
This is illustrated numerically for an ellipsoid  in the next section.

\section{Numerical Results and Discussion} \label{s3}

The surface $\calS$ of the
  ellipsoid      is specified by the position vector
\begin{eqnarray}
&&
\nonumber
\#r_s(\theta,\phi)=c\=U\. \left[\left(\ux\cos\phi+\uy\sin\phi\right)\sin\theta +\uz\cos\theta\right]\,,
\\[5pt]
&&\quad
\quad \theta\in[0,\pi]\,,\quad\phi\in[0,2\pi)\,,
\end{eqnarray}
where
\begin{equation}
\=U=\left(a\ux\ux+b\uy\uy\right)/c+\uz\uz\,
\end{equation}
is the shape dyadic.
Thus, the ellipsoid   has linear
dimensions $2a, 2b$, and $2c$ along the $x$, $y$, and $z$ axes, respectively.  The shape of the
ellipsoid is adequately described by the ratios $a/c$ and $b/c$. After calculating the transition matrix
we
determined the differential scattering, total scattering,
backscattering, and forward-scattering efficiencies in relation to  $\#\Gamma$ and $\koc$, 
while keeping the ratios $a/c\ne 1$ and $b/c\ne 1$ fixed.

The calculation of the transition matrix was accomplished on the Mathematica\texttrademark~platform. 
We used the Gauss--Legendre quadrature scheme
\cite{Jaluria,Sadiku} to evaluate the elements of the matrices $[Y^{(1)}]$ and $[Y^{(3)}]$. By testing against known integrals \cite{GS},  the numbers of nodes for integration over $\theta$ and $\phi$ were chosen to compute the integrals correct to $\pm0.1\%$ relative error.
We used the lower-upper decomposition  method to  invert $[Y^{(1)}]$ \cite{Kreyszig,Jin,Garg}.
The value of $N$
was incremented by unity until 
$\Qb$ converged within a tolerance of $\pm0.1\%$.
 The larger the value of $\Gamma$, the higher was the value
 of $N$ required to achieve convergence.  The highest value of $N$ is $7$ for all results reported here.
 The Mathematica program was verified by comparing with results available for scattering by a sphere ($a=b=c$) made of 
 the chosen Lorentz-nonreciprocal medium \cite{Jafri2}. All results were in complete agreement.

 Next, we present numerical results on   $\Qsca$, $\Qb$, $\Qf$, and $\QD$ as functions of
 \begin{itemize}
\item  the  propagation direction  of the incident plane wave ($\k_{\text{inc}}$),
\item the polarization state of the incident plane wave ($\e_{\text{inc}}$),
\item the magnitude of the magnetoelectric-gyrotropy vector ($\Gamma$),
\item the electrical length of the major axis of the ellipsoid ($\koc$), and
\item  the direction of the magnetoelectric-gyrotropy vector ($\hatGamma$).
\end{itemize}
 Our program accommodates any 
configuration of the incident plane wave and any orientation of the  magnetoelectric-gyrotropy vector. 
 For all of the following results, we set  $a/c=1/2$ and $b/c=2/3$. Comparison
 with the results known for a sphere \cite{Jafri2} helps us identify the shape effects.

\subsection{Effect of $\#\Gamma$}
Let us first focus on the effect of $\#\Gamma$ on $\Qsca$, $\Qb$, and $\Qf$, while keeping the electrical length $\koc$ fixed.

\begin{figure}[htb]
 \centering 
     \begin{subfigure}[h]{0.7\textwidth}
\includegraphics[width=\linewidth]{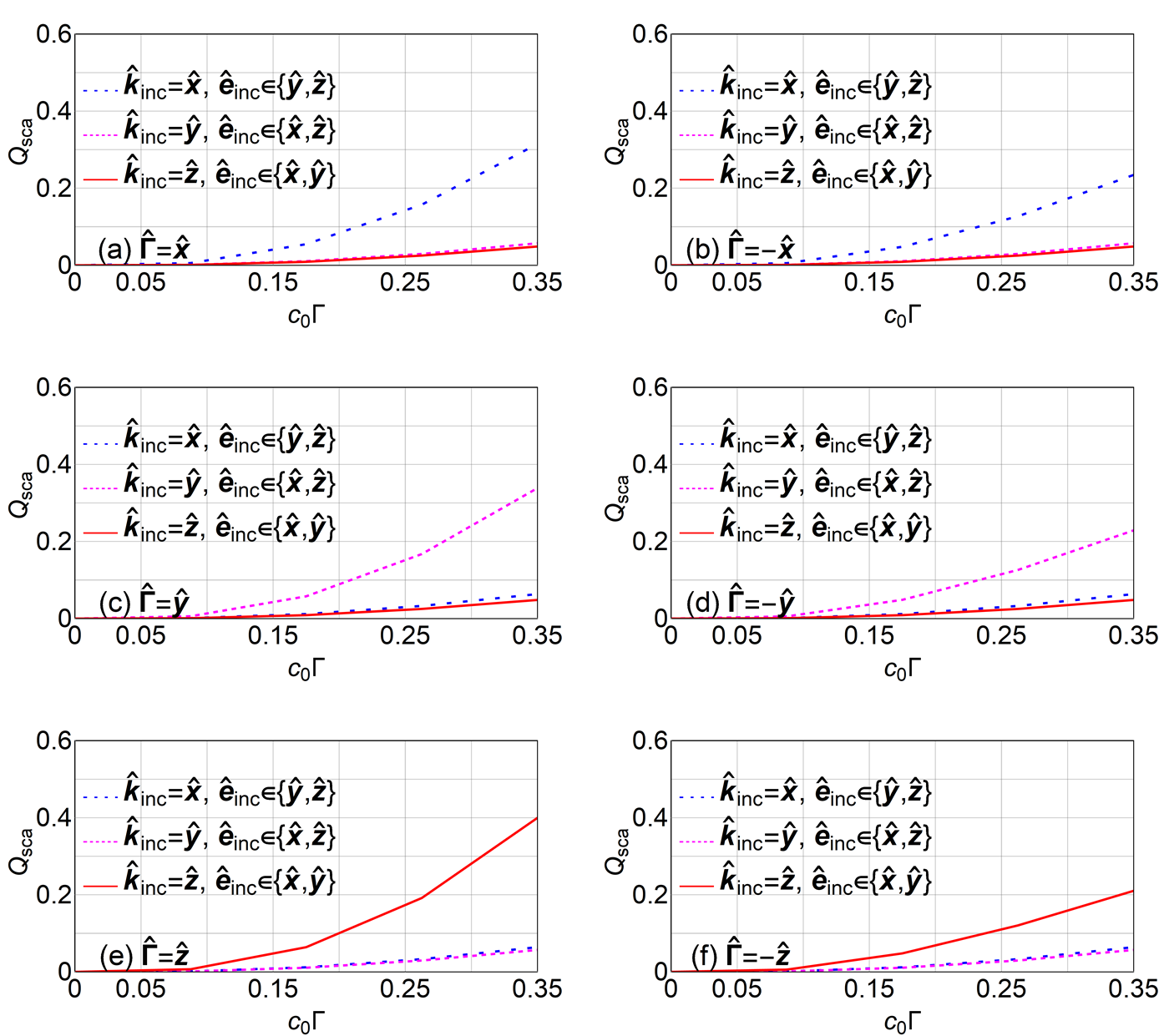}
 \end{subfigure}
\caption{$\Qsca$ vs. $\co\Gamma$  for an ellipsoid  
 composed of
the simplest Lorentz-nonreciprocal medium when $\koc=3$,
$a/c=1/2$, and $b/c=2/3$.
 (a) $\hatGamma =\ux$, (b) $\hatGamma =-\ux$, 
 (c) $\hatGamma =\uy$, (d) $\hatGamma =-\uy$,
(e) $\hatGamma =\uz$, and (f) $\hatGamma =-\uz$.
}
\label{Qs-Gammaxyz}
\end{figure}

Figure~\ref{Qs-Gammaxyz} shows $\Qsca$ vs. $\co\Gamma  \in [0, 0.35 ]$ for an ellipsoid
object composed of
the simplest Lorentz-nonreciprocal medium 
when $\koc=3$. These results were calculated for $\hatGamma \in \{\pm \ux, \pm\uy, \pm\uz \}$, and for all six canonical configurations of the incident plane wave with respect to the semi-axes of the ellipsoid, i.e., 
$\k_{\text{inc}}\in\lec\ux,\uy,\uz\ric$ and $\e_{\text{inc}}\in\lec\ux,\uy,\uz\ric$ such that $\k_{\text{inc}}\perp\e_{\text{inc}}$. 

For any value of $\co\Gamma$ in Fig.~\ref{Qs-Gammaxyz}, $\Qsca$ is   maximum when
$\k_{\text{inc}} = \hatGamma$; furthermore, $\Qsca(-\hatGamma) <\Qsca(\hatGamma)$ then,
as becomes increasingly evident with increase of $\co \Gamma $.  Moreover, $\Qsca$ is highest when $\k_{\text{inc}}$ is parallel to the eigenvector of $\={U}$ 
corresponding to its largest eigenvalue.
 In contrast, when ${\k_{\text{inc}}} \perp \hatGamma$, $\Qsca(\hatGamma)$ is almost indistinguishable from $\Qsca(-\hatGamma)$.

\begin{figure}[htb]
 \centering 
     \begin{subfigure}[h]{0.7\textwidth}
\includegraphics[width=\linewidth]{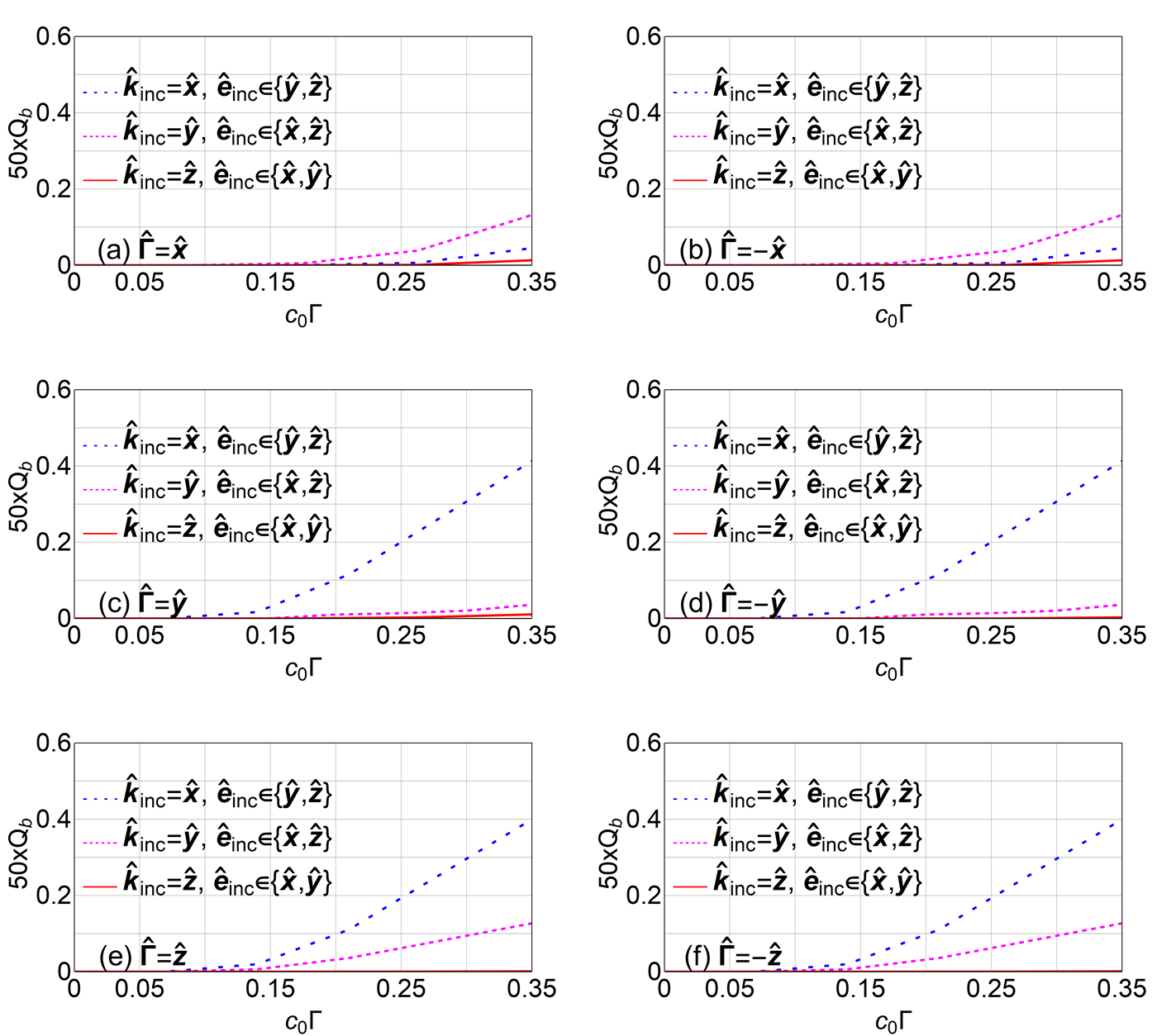}
 \end{subfigure}
\caption{Same as Fig.~\ref{Qs-Gammaxyz}, except for $\Qb$.}
\label{Qb-Gammaxyz}
\end{figure}

Plots of $\Qb$ vs. $\co\Gamma$ are presented in Fig.~\ref{Qb-Gammaxyz}. 
For all values of $\co \Gamma$, when $\k_{\text{inc}} \perp \hatGamma$, $\Qb$ is the smallest
when $\k_{\text{inc}}$ is parallel to the eigenvector of $\={U}$ corresponding to its largest eigenvalue. 
Calculations (not shown) for an  ellipsoid with $a/c=2$ and $b/c=3/2$, for which the eigenvector of $\=U$ corresponding 
to the largest eigenvalue is $\ux$, confirms this conclusion.  
$\Qb$ is much smaller when $\k_{\text{inc}} \parallel \hatGamma$  than when $\k_{\text{inc}} \perp \hatGamma$.
Furthermore, unlike $\Qsca$,  $\Qb(-\hatGamma)$ is almost indistinguishable from $\Qb(\hatGamma)$ 
when $\k_{\text{inc}} \parallel \hatGamma$. 

\begin{figure}[htb]
 \centering 
     \begin{subfigure}[h]{0.7\textwidth}
\includegraphics[width=\linewidth]{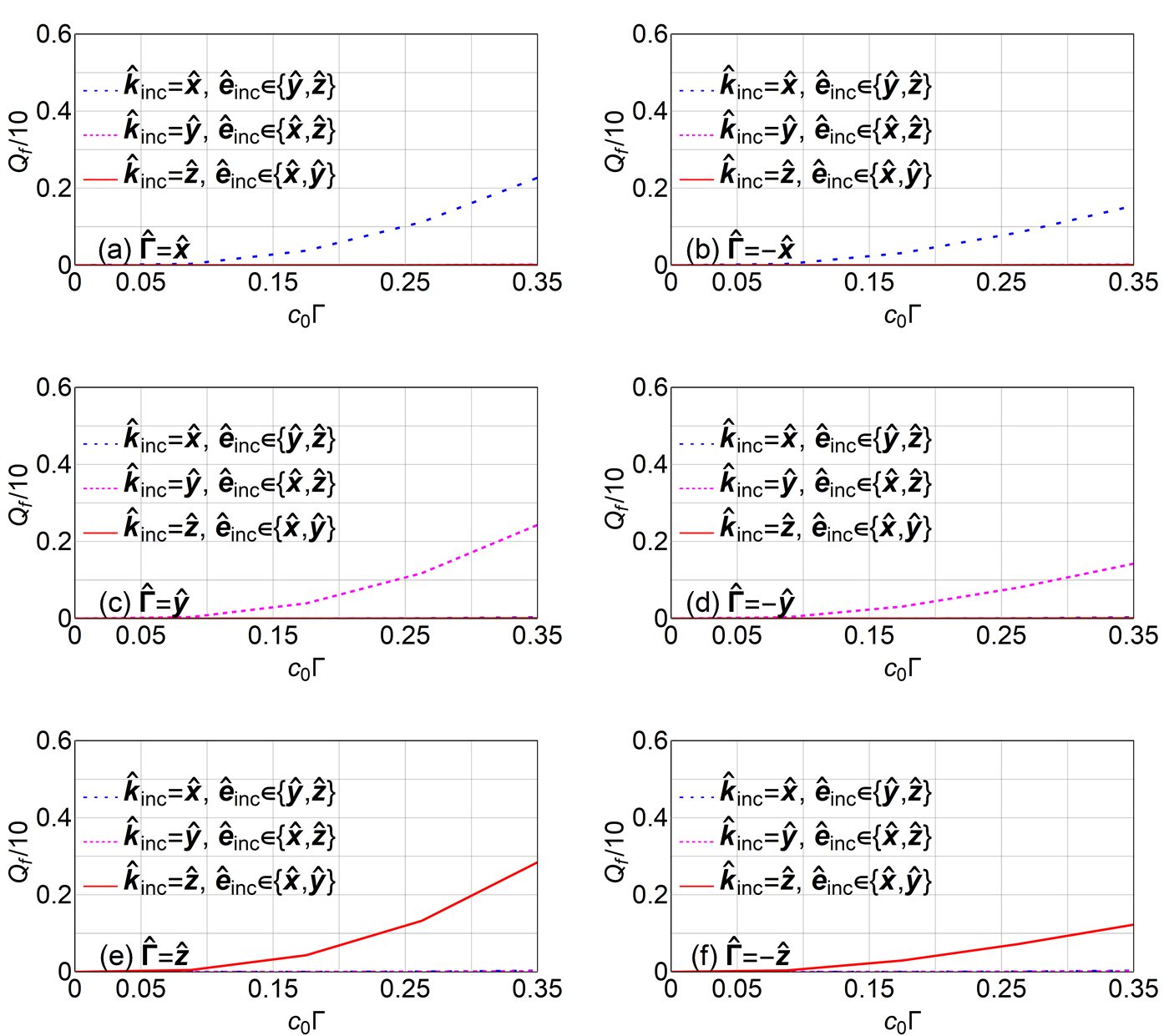}
 \end{subfigure}
\caption{Same as Fig.~\ref{Qs-Gammaxyz}, except for $\Qf$.}
\label{Qf-Gammaxyz}
\end{figure}

Finally, plots of $\Qf$ vs. $\co\Gamma$ are presented in Fig.~\ref{Qf-Gammaxyz}. $\Qf$ is   maximum in this figure when
$\k_{\text{inc}} = \hatGamma$; furthermore, $\Qf(-\hatGamma) <\Qf(\hatGamma)$ then. 
Moreover, $\Qf$ is   maximum when $\k_{\text{inc}}$ is parallel to the eigenvector of $\={U}$ corresponding to its largest eigenvalue,
when $\hatGamma \in \{ \ux, \uy, \uz \}$; however, the direction of $\k_{\text{inc}}$ relative
to the eigenvectors of $\=U$ is virtually inconsequential  when $\hatGamma \in \{ -\ux, -\uy, -\uz \}$.

Let us also note that all scattering efficiencies in Figs.~\ref{Qs-Gammaxyz}--\ref{Qf-Gammaxyz} increase with   $\co \Gamma$. 
Also, the polarization-state-invariance of all scattering efficiencies is evident in these figures.

\subsection{Effect of $\koc$} 

Next, we focus on the effect of $\koc$ on $\QD(\theta,\phi)$, $\Qsca$, $\Qb$, and $\Qf$, 
while keeping the magnitude $\Gamma$ of the magnetoelectric-gyrotropy vector $\#\Gamma$ fixed.

\begin{figure}[h]
 \centering 
     \begin{subfigure}[h]{0.6\textwidth}
\includegraphics[width=\linewidth]{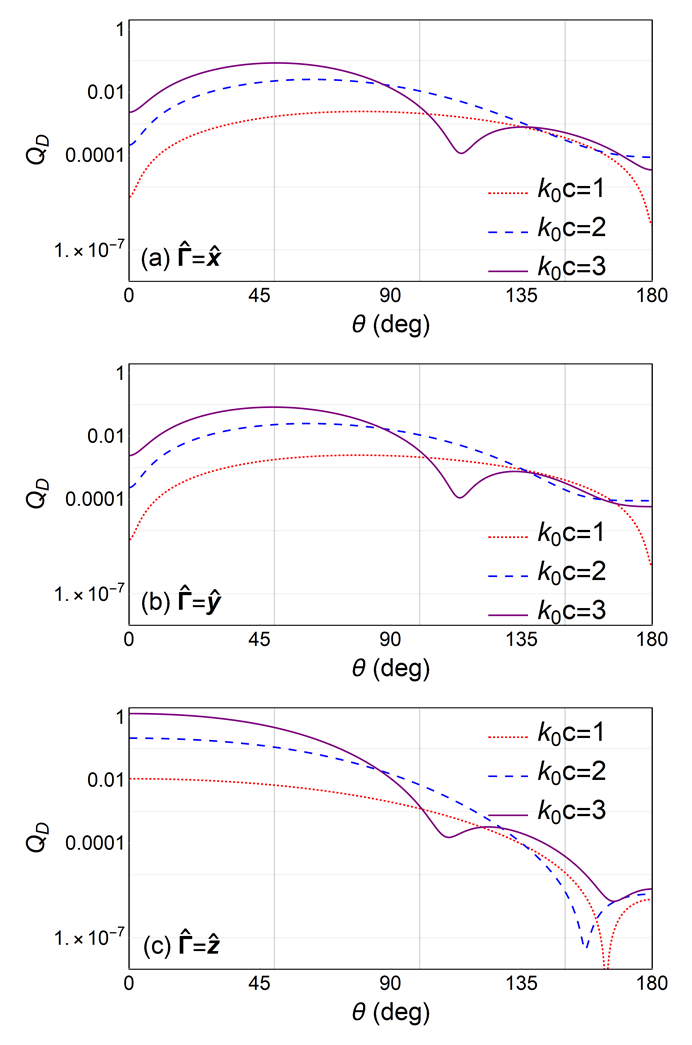}
 \end{subfigure}
\caption{$\QD(\theta,0^\circ)$ vs. $\theta$  for 
an ellipsoid  
 composed of
the simplest Lorentz-nonreciprocal medium when $\koc=3$,
$a/c=1/2$,  $b/c=2/3$, $\co \Gamma=0.25$, $\k_{\text{inc}}=\uz$, and
$\e_{\text{inc}} \in \{ \ux, \uy \}$. The red dotted lines represent $\koc=1$, the blue dashed lines  $\koc=2$, and the purple solid lines  $\koc=3$.  (a) $\hatGamma =\ux$, (b) $\hatGamma =\uy$, (c) $\hatGamma =\uz$.}
\label{QD-Gammaxyz}
\end{figure}

The differential scattering efficiency $\QD(\theta,\phi)$ is shown in Fig.~\ref{QD-Gammaxyz} as a function of $\theta$  for 
$\phi=0^\circ$ when $\k_{\text{inc}}=\uz$,
$\e_{\text{inc}} \in \{ \ux, \uy \}$, and $\hatGamma \in \{\ux, \uy, \uz \}$.   Recall that
$\QD(\theta,\phi)$ has to be independent of $\e_{\text{inc}}$, in accord with Sec.~\ref{sourceinv-plw}. Most
importantly, as $\koc$ increases,  lobes form in the curves of $\QD(\theta,0^\circ)$ regardless of $\hatGamma$,
the same conclusion emerging from similar curves (not shown) for other values of $\phi$.

\begin{figure}[htb]
 \centering 
     \begin{subfigure}[h]{0.6\textwidth}
\includegraphics[width=\linewidth]{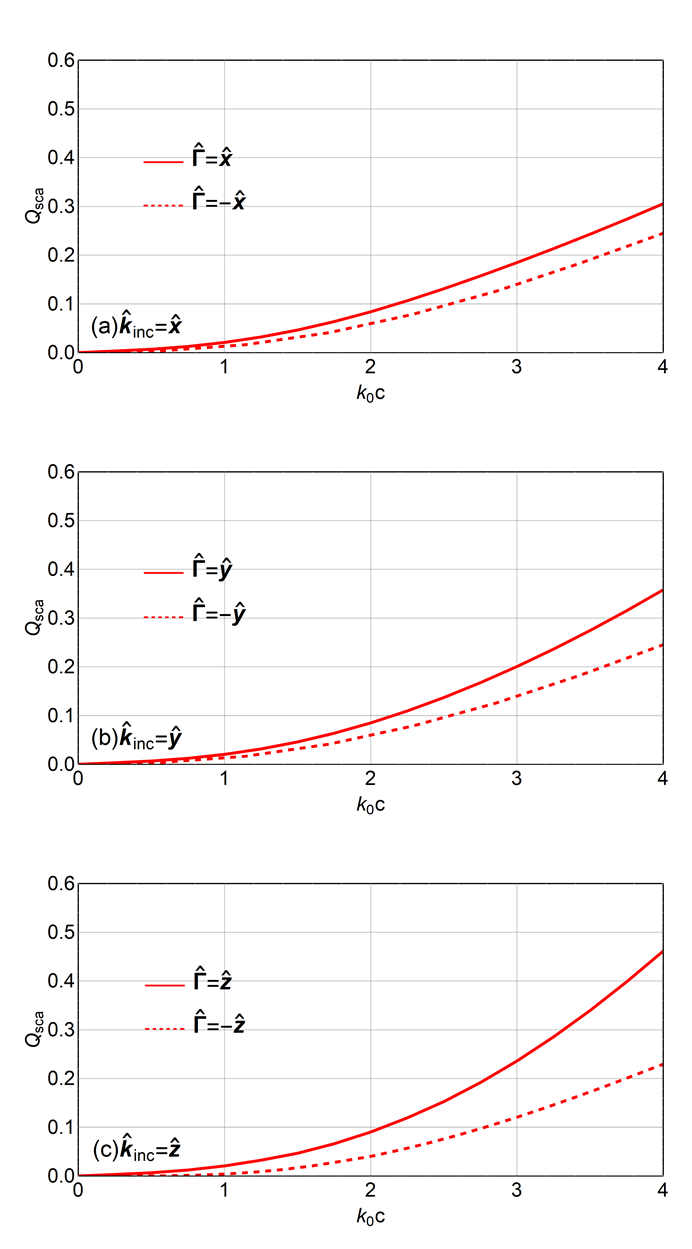}
 \end{subfigure}
\caption{$\Qsca$ vs. $\koc$ for  an ellipsoid  
 composed of
the simplest Lorentz-nonreciprocal medium when $\koc=3$,
$a/c=1/2$,  $b/c=2/3$, $\co \Gamma=0.25$, and  $\hatGamma= \pm\k_{\text{inc}}$.  (a) $\k_{\text{inc}} =\ux$, (b) $\k_{\text{inc}} =\uy$, (c)  $\k_{\text{inc}} =\uz$. }
\label{QS-Gammaz}
\end{figure}

Figure~\ref{QS-Gammaz} shows $\Qsca$ as a function of $\koc$
for an ellipsoid composed
of the simplest Lorentz-nonreciprocal medium with $\co \Gamma=0.25$,
when $\hatGamma= \pm\k_{\text{inc}}$ and
$\k_{\text{inc}}\in\lec \ux, \uy, \uz\ric$. As $\koc$ increases, the excess of
$\Qsca$ for $\hatGamma =\k_{\text{inc}}$ over
$\Qsca$ for $\hatGamma =-\k_{\text{inc}}$ becomes evident. 
Furthermore, this excess is maximum (minimum) when $\hatGamma$ is parallel to the eigenvector of $\={U}$ corresponding to the largest (smallest) eigenvalue. 
The same observations were made for $\Qf$ (not shown). The excess of $\Qb$ for $\hatGamma =\k_{\text{inc}}$ over $\Qb$ for $\hatGamma =-\k_{\text{inc}}$ was very tiny, however (not shown).

\begin{figure}[htb]
 \centering 
     \begin{subfigure}[h]{0.6\textwidth}
\includegraphics[width=\linewidth]{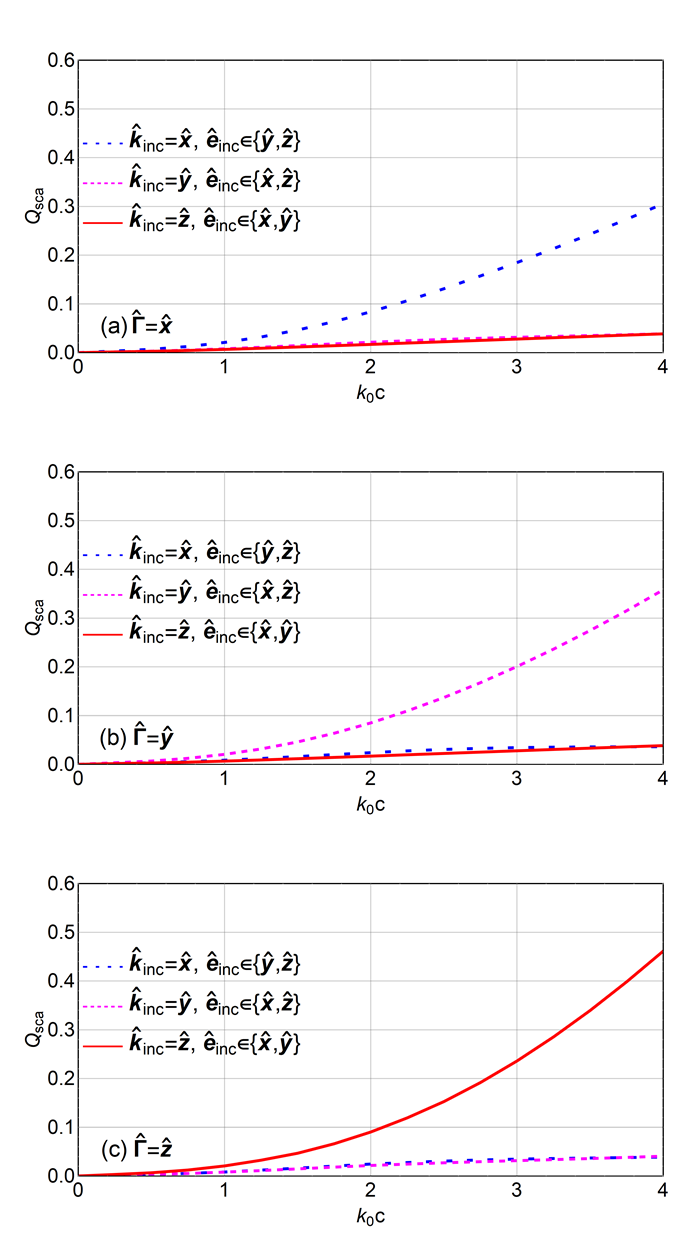}
 \end{subfigure}
\caption{$\Qsca$ vs. $\koc$  for  an ellipsoid  
 composed of
the simplest Lorentz-nonreciprocal medium when $\koc=3$,
$a/c=1/2$,   $b/c=2/3$, $\co \Gamma=0.25$, $\k_{\text{inc}}\in\lec\ux,\uy,\uz\ric$,
and $\e_{\text{inc}}\.\k_{\text{inc}}=0$.
 (a) $\hatGamma =\ux$, (b) $\hatGamma =\uy$, (c) $\hatGamma =\uz$. }
\label{QS-K-E}
\end{figure}

Figure \ref{QS-K-E} shows plots of $\Qsca$ vs. $\koc$ for an ellipsoid composed
of the simplest Lorentz-nonreciprocal medium with $\co \Gamma=0.25$
 when $\k_{\text{inc}}\in\lec\ux,\uy,\uz\ric$ and  $\hatGamma\in\lec\ux,\uy,\uz\ric$.
For any value of $\koc$, $\Qsca$ is highest when $ \hatGamma=\k_{\text{inc}} $; furthermore,
$\Qsca$ is less for $ \hatGamma=-\k_{\text{inc}} $, which follows from Fig.~\ref{QS-Gammaz}.
Additionally, $\Qsca$ is  maximum when $\k_{\text{inc}}$ is parallel to the eigenvector of $\={U}$
corresponding to its largest eigenvalue.

\begin{figure}[htb]
 \centering 
     \begin{subfigure}[h]{0.6\textwidth}
\includegraphics[width=\linewidth]{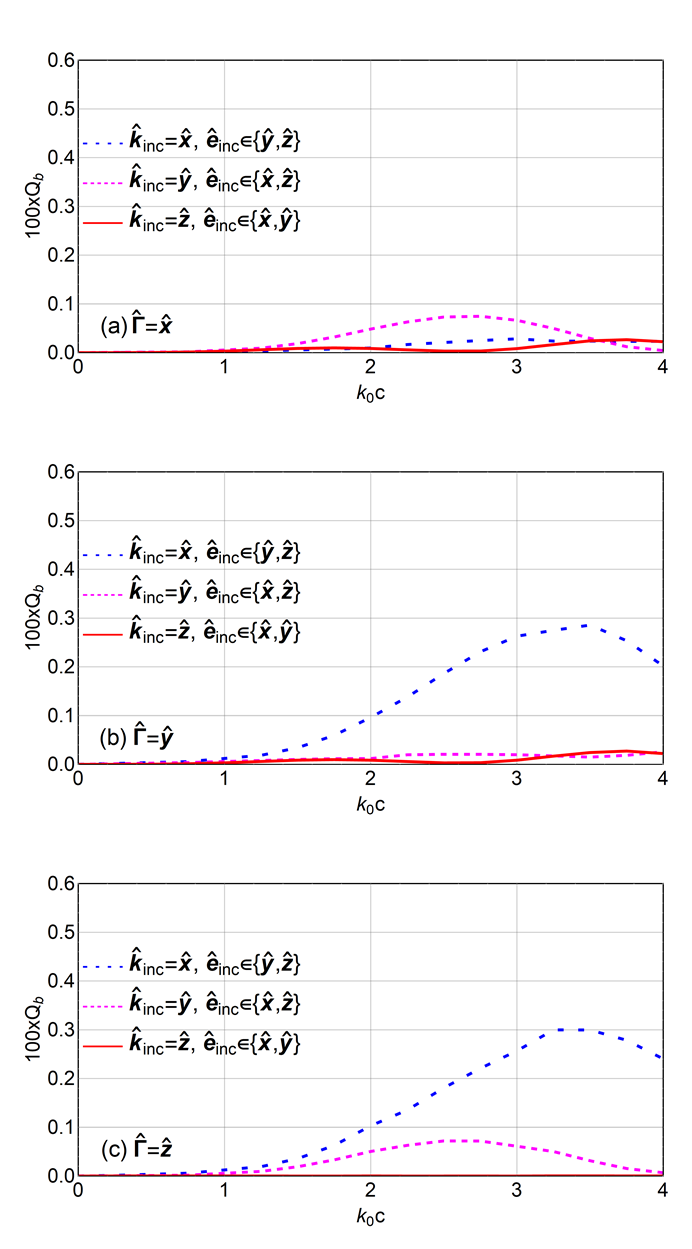}
 \end{subfigure}
\caption{Same as Fig.~\ref{QS-K-E}, except for $\Qb$. }
\label{Qb-Gamma}
\end{figure}

Figure \ref{Qb-Gamma} shows $\Qb$ vs. $\koc$ for an ellipsoid composed
of the simplest Lorentz-nonreciprocal medium with $\co \Gamma=0.25$
when $\k_{\text{inc}}\in\lec\ux,\uy,\uz\ric$ and  $\hatGamma\in\lec\ux,\uy,\uz\ric$. As mentioned in Sec.~\ref{s3}.A,  $\Qb(-\hatGamma)$ is almost indistinguishable from $\Qb(\hatGamma)$ 
when $\k_{\text{inc}} \parallel \hatGamma$. 
We conclude from Fig.~\ref{Qb-Gamma} that
$\Qb$ is the minimum when
either $ \hatGamma=\k_{\text{inc}}$ or $ \hatGamma=-\k_{\text{inc}}$, regardless of $\koc$, for a fixed $\k_{\text{inc}}$. In this trend, $\Qb$ differs from $\Qsca$.

\begin{figure}[htb]
 \centering 
     \begin{subfigure}[h]{0.6\textwidth}
\includegraphics[width=\linewidth]{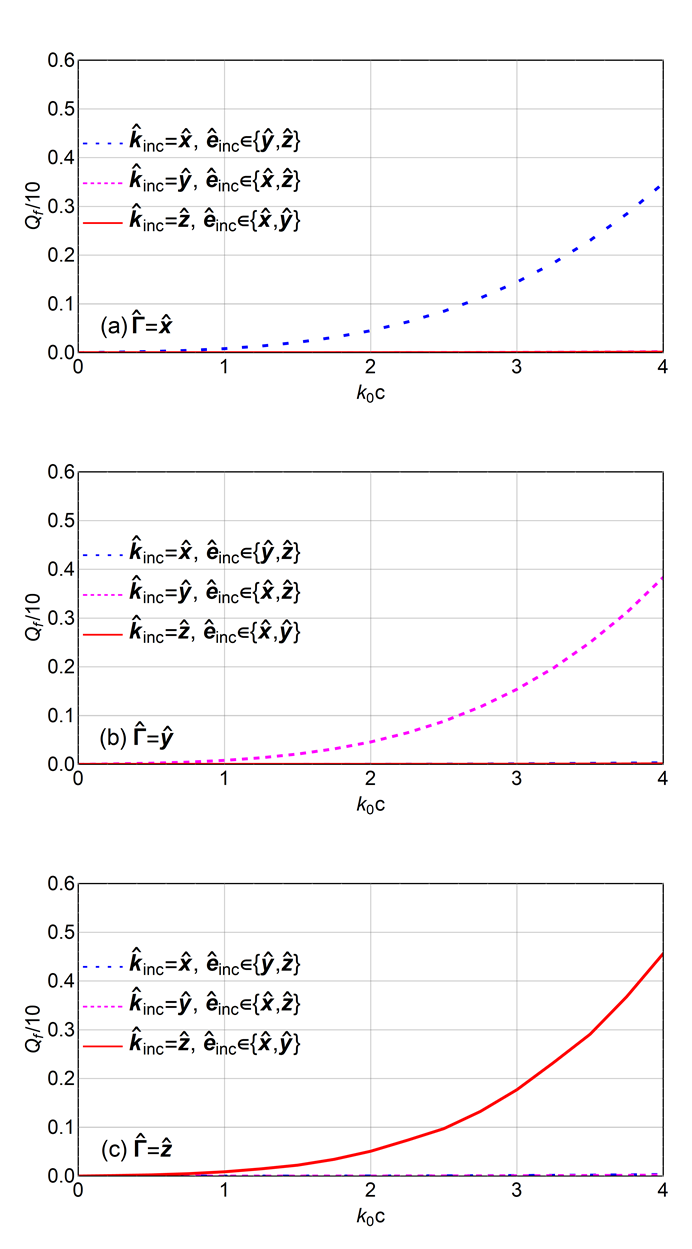}
 \end{subfigure}
\caption{Same as Fig.~\ref{QS-K-E}, except for $\Qf$.}
\label{Qf-Gamma}
\end{figure}

 Figure \ref{Qf-Gamma} is the counterpart of Fig.~\ref{Qb-Gamma}
for $\Qf$. Like $\Qsca$, $\Qf$ is   maximum when
$\hatGamma=\k_{\text{inc}}$, and is less when $\hatGamma=-\k_{\text{inc}}$ based on an argument similar to that of Fig. \ref{QS-Gammaz} but for $\Qf$ (not shown). Moreover, $\Qf$ is  maximum when $\k_{\text{inc}}$ is parallel to the eigenvector of $\={U}$ corresponding to its largest eigenvalue.

\section{Concluding Remarks}\label{s4}
The  medium described by Eqs.~(\ref{conrel0}) 
 is the  simplest  Lorentz-nonreciprocal medium, differing from the   free space by virtue of a non-null magnetoelectric-gyrotropy
 vector $\#\Gamma$. When an object composed of this medium
suspended in   free space is irradiated by electromagnetic fields
 from a source, the differential scattering efficiency is immune to the transformation of the incident toroidal electric field phasor  into a poloidal electric field phasor, or \textit{vice versa}. Of course, the magnetic field phasor accompanying a toroidal/poloidal electric field phasor is poloidal/toroidal, by virtue of the Faraday equation. This toroidal-poloidal source-invariance is also exhibited by the total scattering efficiency.
 
A consequence of the toroidal-poloidal source-invariance of the
differential scattering efficiency is the polarization-state invariance of
the  differential scattering efficiency when the irradiating field is a plane wave. As a result, not only the total scattering efficiency but also the forward-scattering  and the backscattering efficiencies also
exhibit polarization-state invariance.

Numerical results  obtained using the extended boundary condition method for plane-wave scattering by an
ellipsoid composed of the simplest Lorentz-nonreciprocal medium validated the foregoing conclusions. Furthermore,
regardless of the magnitude of the magnetoelectric-gyrotropy vector and the electrical size of the ellipsoid, the total scattering and forward-scattering efficiencies are maximum when the plane wave is incident in a direction that  is  {coparallel (but not antiparallel)} to the magnetoelectric-gyrotropy vector, as compared to when the incidence direction is antiparallel or perpendicular to the magnetoelectric-gyrotropy vector. The backscattering efficiency is minimum when the magnetoelectric-gyrotropy vector  is  {parallel} to the 
incidence  direction.

When the incidence direction is parallel to the eigenvector of the ellipsoid's shape dyadic corresponding to its largest eigenvalue, the total scattering and the forward-scattering efficiencies are maximum, provided that the incidence direction is coparallel (but not antiparallel) to the magnetoelectric-gyrotropy vector.

As the electrical size of the ellipsoid increases, lobes appear in the curves of the differential scattering efficiency, regardless of the direction of the magnetoelectric-gyrotropy vector. Furthermore, the excess of the total scattering and forward-scattering efficiencies when the magnetoelectric-gyrotropic vector is  {coparallel (but not antiparallel)} to the incidence direction over when it is antiparallel, increases as the electrical size of the ellipsoid increases. Maximum excess is achieved when the magnetoelectric-gyrotropic vector is parallel to the eigenvector of the shape dyadic corresponding to its largest eigenvalue. Thus, the simplest manifestation of Lorentz nonreciprocity in an object is intimately connected to the shape of that object in affecting the scattered field.

\vspace{0.5cm}
\noindent {\bf Acknowledgment.}  AL thanks the Charles Godfrey Binder Endowment at Penn State for ongoing support of his research
activities.

\end{document}